\documentclass[aps,preprint,showpacs,preprintnumbers,amsmath,amssymb]{revtex4}

\usepackage{graphics,epsfig}
\usepackage{graphicx}
\usepackage{dcolumn}
\usepackage{bm}
\usepackage{float}

\begin{document}

\title{Quasinormal modes and second order thermodynamic phase
transition for Reissner-Nordstr\"om black hole}
\author{Jiliang Jing} \email{jljing@hunnu.edu.cn} \author{Qiyuan Pan}
\affiliation{ Institute of Physics and  Department of Physics, \\
Hunan Normal University, Changsha, Hunan 410081, P. R. China
 \\ and
\\
Key Laboratory of Low Dimensional Quantum Structures and Quantum
Control (Hunan Normal University), Ministry of Education, P. R.
China}

\vspace*{0.2cm}
\begin{abstract}
\vspace*{0.2cm}

The relation between the quasinormal modes (QNMs) and the second
order thermodynamic phase transition (SOTPT) for the
Reissner-Nordstr\"om (RN) black hole is studied. It is shown that
the quasinormal frequencies of the RN black hole start to get a
spiral-like shape in the complex $\omega$ plane and both the real
and imaginary parts become the oscillatory functions of the charge
if the real part of the quasinormal frequencies arrives at its
maximum at the second order phase transition point of Davies for
given overtone number and angular quantum number. That is to say, we
can find out the  SOTPT  point from the QNMs of the RN black hole.
The fact shows that the quasinormal frequencies carry the
thermodynamical information of the RN black hole.

\end{abstract}


 \vspace*{1.5cm}

\maketitle


The QNMs of a black hole are defined as proper solutions of the
perturbation equations belonging to certain complex characteristic
frequencies which satisfy the boundary conditions appropriate for
purely ingoing waves at the event horizon and purely outgoing waves
at infinity \cite{Chand75}. They are entirely fixed by the structure
of the background spacetime and irrelevant of the initial
perturbations \cite{Chand75, Andersson}. Thus, it is generally
believed that QNMs carry the footprint to directly identify the
existence of a black hole. Meanwhile, the study of QNMs may lead to
a deeper understanding of the thermodynamical properties of black
holes in loop quantum gravity \cite{Hod} \cite{Dreyer}, as well as
the QNMs of anti-de Sitter black holes have a direct interpretation
in terms of the dual conformal field theory \cite{Maldacena, Witten,
Kalyana}.

On the other hand, one of important characteristics of a black hole
is its thermodynamical properties. It is well known that the heat
capacity of the Schwarzschild black hole is always negative and so
the black hole is thermodynamical unstable. But for the RN black
hole, the heat capacity is negative in some parameter region and
positive in other region. Davies pointed out that the phase
transition appears in black hole thermodynamics and the SOTPT takes
place at the point where the heat capacity diverges \cite{Davies1,
Davies2, Davies3}.

Because the QNMs of a black hole are entirely fixed by the structure
of the background spacetime and the SOTPT is only related to the
parameters of the black hole, an interesting question is whether
there are some relations between them. The aim of this paper is to
study this question for the RN black hole, and we find that the
quasinormal frequencies do carry the thermodynamical information of
the black hole.

Assuming that the azimuthal and time dependence of the fields will
be the form $e^{-i(\omega t-m\varphi)}$ and using the Newman-Penrose
formalism \cite{Newman}, we can obtain the separated equations for
massless scalar,  Dirac, and Rarita-Schwinger (RS) perturbations
around a  RN  black hole \cite{Jing, Khanal, Castillo}
\begin{eqnarray}\label{w4}
&&[\Delta {\mathcal{D}}_{1-s}{\mathcal{D}}_{0}^{\dag}
+2(2s-1)i\omega r-(\lambda +2s)]\Delta^{s}R_{s} =0,~~~\nonumber \\
&&[{\mathcal{L}}_{1-s}^{\dag}{\mathcal{L}}_{s}+(\lambda
+2s)]S_{s}=0,~~\text{(s=0,~ 1/2,~ 3/2)}~~~~\\ \nonumber \\ \nonumber
\label{w5}
 &&[\Delta
{\mathcal{D}}_{1+s}^{\dag}{\mathcal{D}}_{0}
+2(2s+1)i\omega r-\lambda  ]R_{s} =0,~~~~~~~~~~~~~~~~\nonumber \\
&&({\mathcal{L}}_{1+s}{\mathcal{L}}_{-s}^{\dag}+\lambda  )S_{s}=0,
~~~~\text{(s=0,~ -1/2,~ -3/2)}
\end{eqnarray}
where $\Delta=r^2-2Mr+Q^2$, $M$ and $Q$ represent the mass and
charge of the black hole, and $\lambda  $ is the angular separation
constant \cite{WS,ET,ET-1}
 \begin{eqnarray}
 \label{As}
\lambda  = \left\{
\begin{array}{ll} (l-s)(l+s+1),~~~~~ l=|s|,|s|+1,\cdots, \\
 (j-s)(j+s+1),~~~~ j=|s|,|s|+1,\cdots,
\end{array} \right.
 \end{eqnarray}
where $l$ and $j$  are the quantum number characterizing the angular
distribution for the boson and fermion perturbations respectively.
Introducing an usual tortoise coordinate $ dr_*=r^2/\Delta  dr $ and
resolving the equation in the form $ R_{s}=\Delta^{-s/2}\Psi_{s}/r,
$ we can rewrite the radial wave equations in  Eqs. (\ref{w4}) and
(\ref{w5}) as
\begin{eqnarray}\label{wave}
\frac{d^{2}\Psi_{s} }{d r_{*}^2}+[\omega ^2-V]\Psi_{s} =0,
\end{eqnarray}
with
\begin{eqnarray}\label{Poten}
V= i s \omega r^2\frac{d}{d r}\frac{\Delta }{r^4}+\frac{(s+\lambda
)\Delta +\left(\frac{s}{2}\frac{d\Delta }{dr}\right)^{2}}{r^4}
+\frac{\Delta }{r^3}\frac{d}{dr} \frac{\Delta}{r^2}.
\end{eqnarray}

The boundary conditions on wave function $\Psi_s$ at the horizon and
infinity can be expressed as
 \begin{eqnarray}
 \label{Bon}
\Psi_s  \sim \left\{
\begin{array}{ll} (r-r_+)^{-\frac{s}{2}-\frac{i\omega}{2\kappa_{+}}} &
~~~~r\rightarrow r_+, \\
     r^{-s+i\omega}e^{i\omega r} & ~~~~     r\rightarrow +\infty,
\end{array} \right.
 \end{eqnarray}
where $\kappa_\pm=(r_+-r_-)/(2r^2_\pm)$ is the surface gravity on
the horizons $r_\pm$. A solution to Eq. (\ref{wave}) that has the
desired behavior at the boundary can be written as
 \begin{eqnarray}\label{expand}
 \Psi_s&=&r(r-r_+)^{-\frac{s}{2}
 -\frac{i\omega}{2\kappa_+}}(r-r_-)^{-1-\frac{s}{2}+2i\omega+
 \frac{i\omega}{2\kappa_-}}\nonumber\\
 &&\times e^{i\omega (r-r_-)}\sum_{m=0}^{\infty}
 a_m\left(\frac{r-r_+}{r-r_-}\right)^m.
 \end{eqnarray}
If we take $r_++r_-=1$, the sequence of the expansion coefficients
$\{a_m: m=1,2,....\}$ is determined by a three-term recurrence
relation staring with $a_0=1$:
 \begin{eqnarray} \label{rec}
 &&\alpha_0 a_1+\beta_0 a_0=0, \nonumber \\
 &&\alpha_m a_{m+1}+\beta_m a_m+\gamma_m a_{m-1}=0,~~~m=1,2,....
 \end{eqnarray}
The recurrence coefficient $\alpha_m$, $\beta_m$ and $\gamma_m$ are
given in terms of $m$ and the black hole parameters by
\begin{eqnarray}
 &&\alpha_m=m^2+(C_0+1)m+C_0, \nonumber \\
 &&\beta_m=-2m^2+(C_1+2)m+C_3, \nonumber  \\
 &&\gamma_m=m^2+(C_2-3)m+C_4-C_2+2,
 \end{eqnarray}
and the intermediate constants $C_m$ are defined by
\begin{eqnarray}
 C_0&=&1-s-i\omega-i\omega B, \nonumber \\
 C_1&=&-4+2i\omega(2+b)+2i\omega B,  \nonumber  \\
 C_2&=&s+3-3i\omega-i\omega B, \nonumber \\
 C_3&=&\omega^2(4+2b-4r_+r_-)-s-1+(2+b)i\omega\nonumber \\&&
 -\lambda^2+(2\omega+i)
 \omega B,
 \nonumber \\
 C_4&=&s+1+2\left(i \omega-s-\frac{3}{2}\right)i\omega-(2\omega+i)\omega B,
 \end{eqnarray}
where $B=(r_+^2+r_-^2)/(r_+-r_-)$. The series in (\ref{expand})
converges and the $r=+\infty$ boundary condition (\ref{Bon}) is
satisfied if, for a given $s$ and $\lambda$, the frequency $\omega$
is a root of the continued fraction equation
\begin{eqnarray}\label{ann}
 &&\left[\beta_m-\frac{\alpha_{m-1}\gamma_m}{\beta_{m-1}-}
 \frac{\alpha_{m-2}\gamma_{m-1}}{\beta_{m-2}-}...
 \frac{\alpha_0\gamma_1}{\beta_0}\right]\nonumber \\
 &&=
 \left[\frac{\alpha_m\gamma_{m+1}}{\beta_{m+1}-}
 \frac{\alpha_{m+1}\gamma_{m+2}}{\beta_{m+2}-}
 \frac{\alpha_{m+2}\gamma_{m+3}}{\beta_{m+3}-}...\right],
 ~~~~~~~~~~~~~~~~~~~~~~~~(m=1,2...).
 \end{eqnarray}
This leads to a simple method to find quasinormal frequencies of the
RN black hole --- defining a function which returns the value of the
continued fraction for an initial guess at the frequency, and then
use a root finding routine to find the zeros of this function in the
complex $\omega$ plane. The frequency for which happens is a
quasinormal frequency \cite{Leaver,Leaver1}.

\begin{figure}
\includegraphics[scale=1.2]{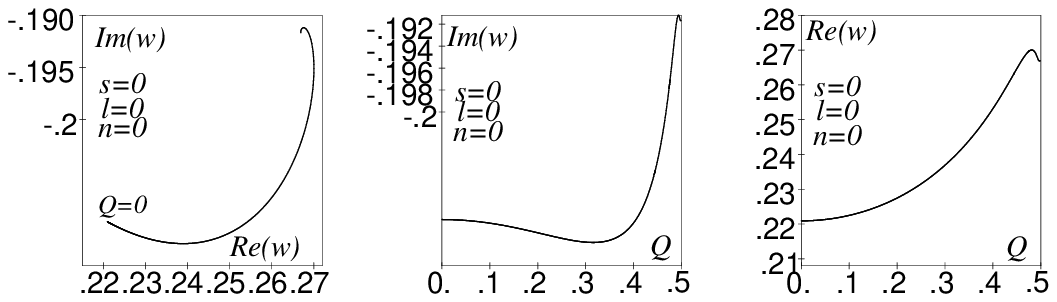} \vspace*{0.10cm} \\
\includegraphics[scale=1.2]{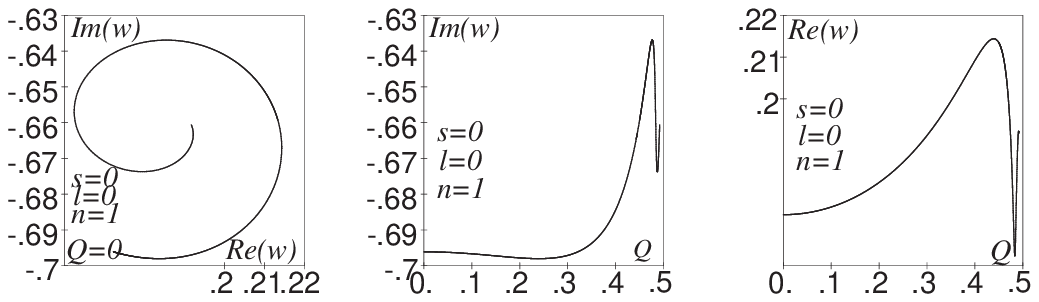} \vspace*{0.10cm} \\
\includegraphics[scale=1.2]{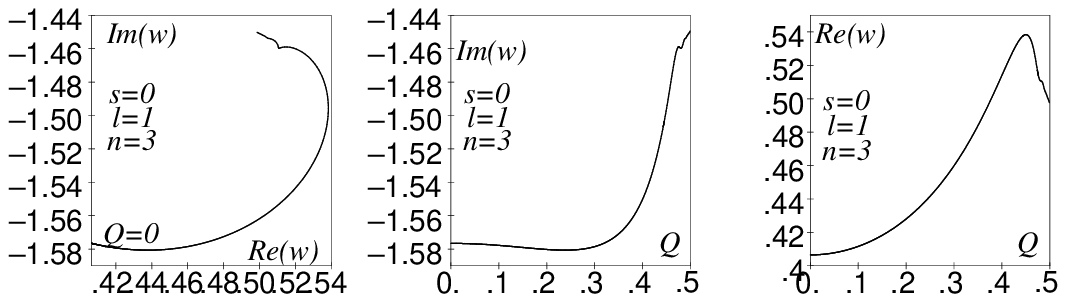} \vspace*{0.10cm} \\
\includegraphics[scale=1.2]{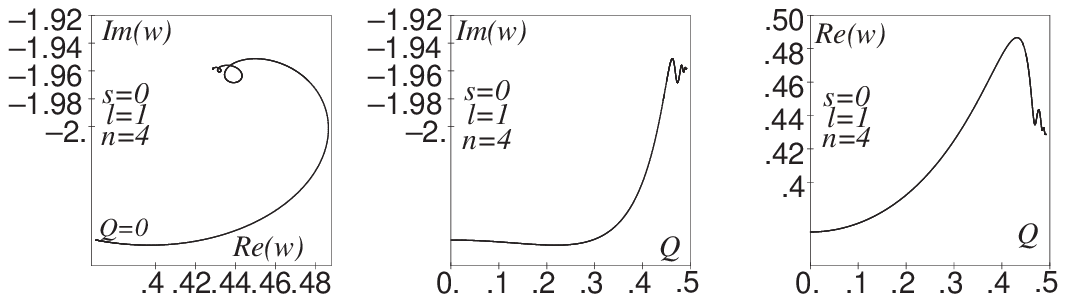}  \\
\caption{\label{fig1} Left four panels show trajectories in the
complex $\omega$  plane of the scalar quasinormal frequencies of the
RN black hole for $l=0,~ n=1,~2 $ and $l=1,~ n=3,~4 $. The other
panels draw the imaginary part $Im(\omega)$ and real part
$Re(\omega)$  of the quasinormal frequencies versus the charge $Q$.
}
\end{figure}

\begin{figure}
\includegraphics[scale=1.2]{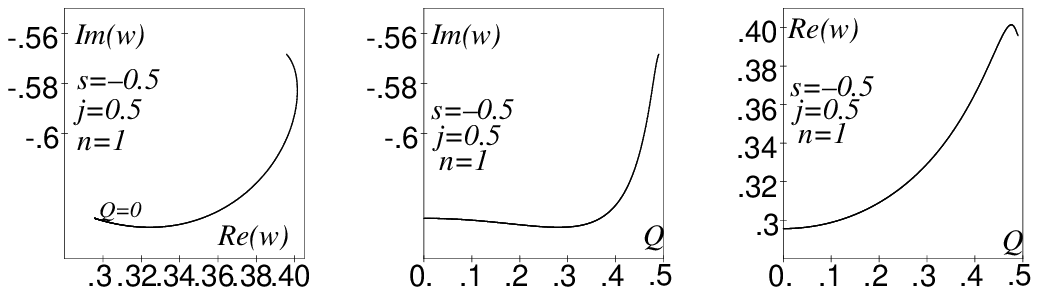} \vspace*{0.10cm} \\
\includegraphics[scale=1.2]{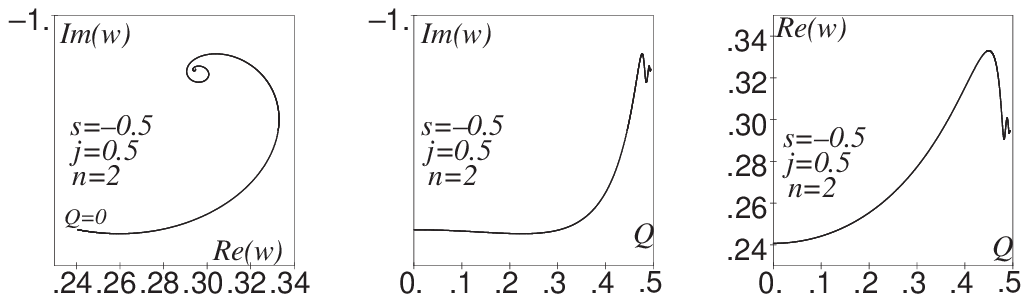} \vspace*{0.10cm} \\
\includegraphics[scale=1.2]{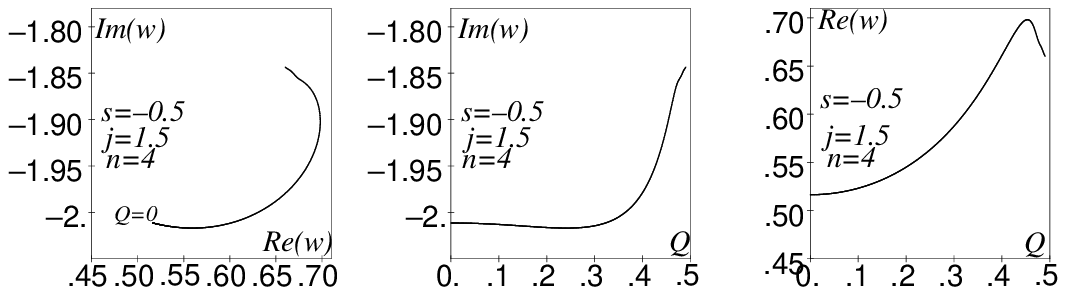} \vspace*{0.10cm} \\
\includegraphics[scale=1.2]{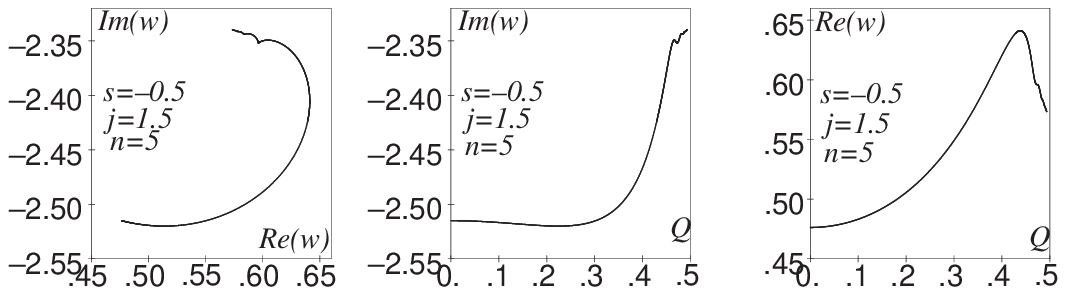}  \\
\caption{\label{fig2} Left four panels show trajectories in the
complex $\omega$  plane of the Dirac quasinormal frequencies  for
$j=1/2,~ n=1,~2 $ and $j=3/2,~ n=4,~5 $. The other panels draw the
imaginary part $Im(\omega)$ and real part $Re(\omega)$  of the
quasinormal frequencies versus the charge $Q$.}
\end{figure}

\begin{figure}
\includegraphics[scale=1.2]{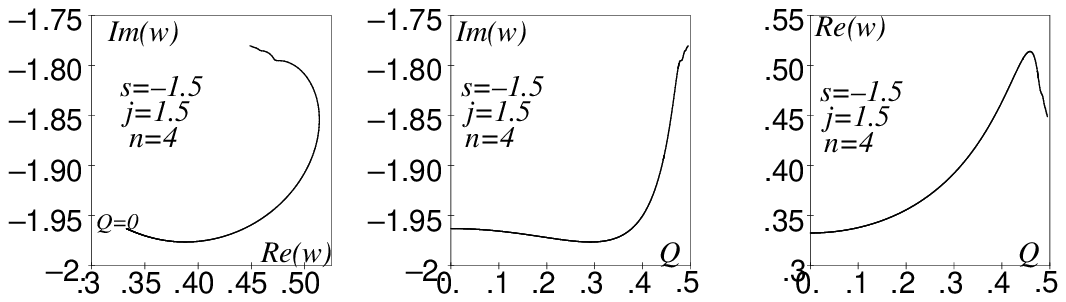} \vspace*{0.10cm} \\
\includegraphics[scale=1.2]{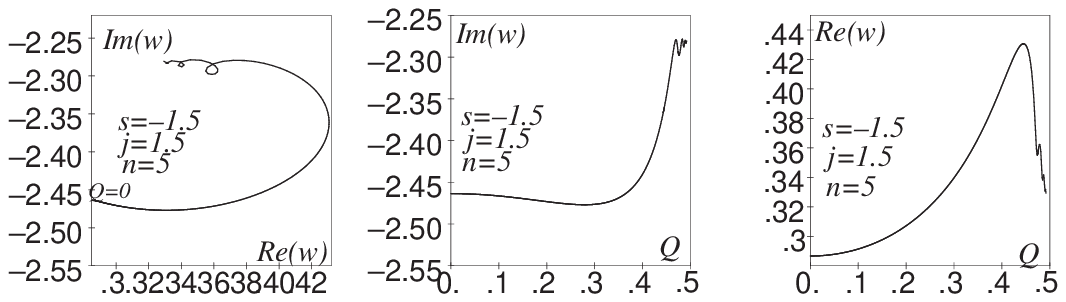} \vspace*{0.10cm} \\
\includegraphics[scale=1.2]{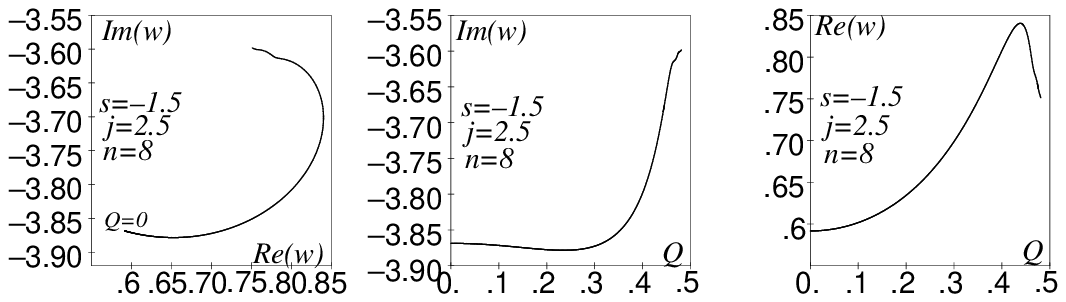} \vspace*{0.10cm} \\
\includegraphics[scale=1.2]{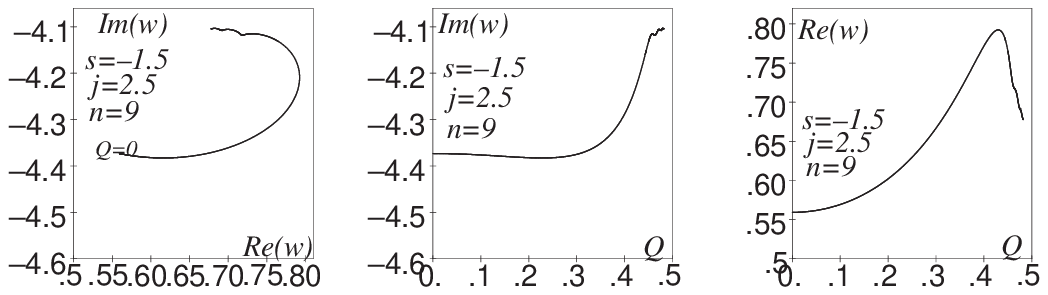}  \\
\caption{\label{fig3} Left four panels show trajectories in the
complex $\omega$  plane of the RS quasinormal frequencies for
$j=3/2,~ n=4,~5 $ and $j=5/2,~ n=8,~9 $. The other panels draw the
imaginary part $Im(\omega)$ and real part $Re(\omega)$  of the
quasinormal frequencies versus the charge $Q$.}
\end{figure}

The Figs.  \ref{fig1}-\ref{fig3} describe the QNMs for the scalar
(s=0),  Dirac (s=-1/2), and RS fields (s=-3/2) obtained by the
continued fraction method respectively. In Fig. \ref{fig1} (or
\ref{fig2}, \ref{fig3}) left four panels show trajectories in the
complex $\omega$  plane of the scalar (or Dirac, RS) quasinormal
frequencies of the RN black hole with different angular quantum
number and overtone number. The other panels draw the imaginary part
$Im(\omega)$ and real part $Re(\omega)$  of the quasinormal
frequencies versus the charge $Q$. These figures tell us that in the
complex $\omega$ plane the quasinormal frequencies will get a
spiral-like shape as the charge $Q$ increases to its extremal value
when the overtone number equals to or exceeds a critical value $n_c$
for a fixed angular quantum number (say, $n_c=1$ with $l=0$ and
$n_c=4$ with $l=1$ for the scalar field,  $n_c=2$ with $j=1/2$ and
$n_c=5$ with $j=3/2$ for the Dirac field, and $n_c=5$ with $j=3/2$
and $n_c=9$ with $j=5/2$ for the RS field.), and at same time both
the real and imaginary parts become the oscillatory functions of the
charge. The critical value of the overtone number $n_c$ increases as
the angular quantum number $l$ (or $j$) increases for a given
perturbation, and it also increases as the absolute value of spins
$|s|$ increases for the same level of the angular quantum number
(here the least level is defined as $l=0$,  $j=1/2$, and $j=3/2$ for
the scalar,  Dirac, and RS fields respectively).

\begin{figure}
\includegraphics[scale=1.6]{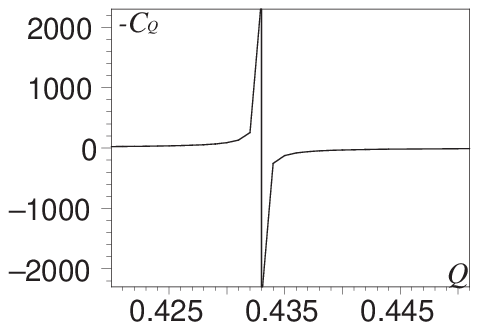}
\caption{\label{fig5} The panel shows trajectory of $-C_Q$ versus
$Q$ near the singular point of the heat capacity $Q_{sp}=0.433$.}
\end{figure}

On the other hand, we know that the heat capacity $C_{Q}$ of the RN
black hole is given by \cite{Davies1,Davies2}
\begin{eqnarray}
&&C_{Q}=T\left(\frac{\partial S}{\partial T}\right)_{Q}
=\frac{TS^{3}M}{\pi Q^{4}/4-S^{3}T^{2}},
\end{eqnarray}
where $T$ is the temperature and $S$ is the entropy of the black
hole. It is obvious that the singular point of the heat capacity
(SPHC) occurs at $\pi Q^{4}/4-S^{3}T^{2}=0$, i.e., $Q_{sp}=\sqrt{3}
M /2 \approx 0.4330127$. Fig. \ref{fig5} shows that the heat
capacity is negative in the region $Q<Q_{sp}$ and positive in the
region $Q>Q_{sp}$. Davies \cite{Davies1, Davies2, Davies3} showed
that the SPHC  is just the SOTPT point of the black hole
thermodynamics.

To find the relation between the QNMs and the SOTPT point, we
present critical point $Q_{cp}$,  which is the value of the charge
at which the real part of the first oscillatory quasinormal
frequencies arrives at its maximum, with the different critical
overtone number $n_c$ for the scalar, Dirac and RS fields in Table
\ref{table1}. From the table we find that the SPHC $Q_{sp}$ is in
good agreement with the critical point $Q_{cp}$ because the
difference between $Q_{sp}$ and $Q_{cp}$ is less than $2.55\%$.
Besides, for the Schwarzschild black hole there is no SPHC because
its heat capacity is always negative and there is no critical point
because its QNMs never get a spiral-like.

\begin{table}[ht]
\caption{\label{table1}  The values of the charge for  $Q_{cp}$. }
\begin{tabular}{c|c|c}
 \hline \hline
 & \multicolumn {2} {c} {Scalar field~~ (s=~0)}
 \\   \hline ~~~~($l$, $n_c$)~~~~   & ~~~~( 0, ~1 ) ~~~~& ~~~~(1,~  4)~~~~
\\    \hline   $Q_{cp}$   & 0.438 & 0.432
 \\   \hline  $\frac{|\Delta Q|}{Q_{cp}}$   & 1.15\%  & 0.23\%
          \\ \hline \hline
 &  \multicolumn {2} {c} {Dirac field ~~  (s=~-1/2)}
 \\ \hline ~($j$, $n_c$)~ & (1/2, ~2 )~ & (3/2,~ 5)
 \\ \hline  $Q_{cp}$ &     0.442 & 0.436
 \\ \hline  $\frac{|\Delta Q|}{Q_{cp}}$ & 2.08\% & 0.69\%
\\  \hline \hline

 &   \multicolumn {2} {c}  {~~Rarita-Schwinger field ~~(s=~-3/2)~~}
 \\  \hline ~($j$, $n_c$)~& (3/2, ~5) ~& (5/2, ~9)
 \\  \hline  $Q_{cp}$  & 0.444 & 0.430
 \\  \hline $\frac{|\Delta Q|}{Q_{cp}}$ & 2.55\% & 0.69\%
\\
 \hline \hline
\end{tabular}
\end{table}

\begin{figure}
\includegraphics[scale=1.2]{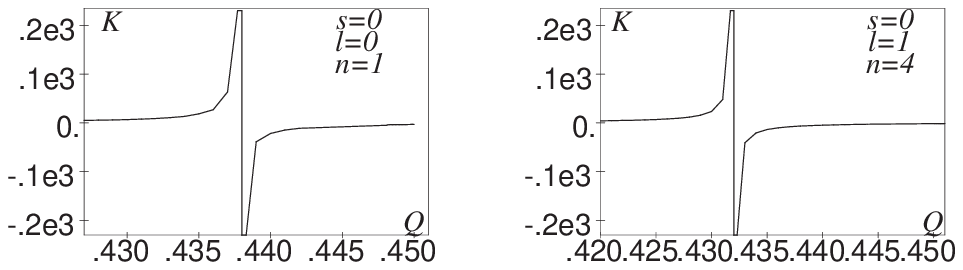}   \\
\includegraphics[scale=1.2]{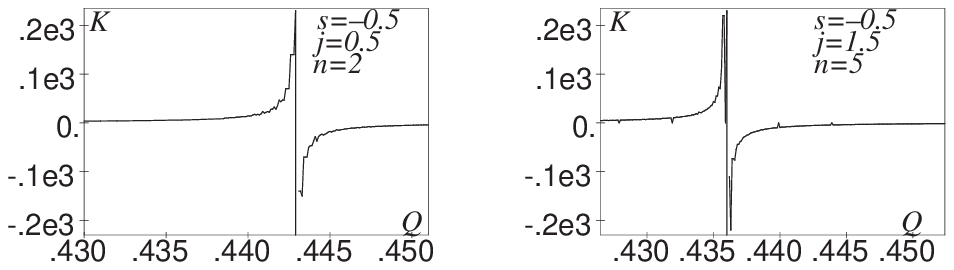}   \\
\includegraphics[scale=1.2]{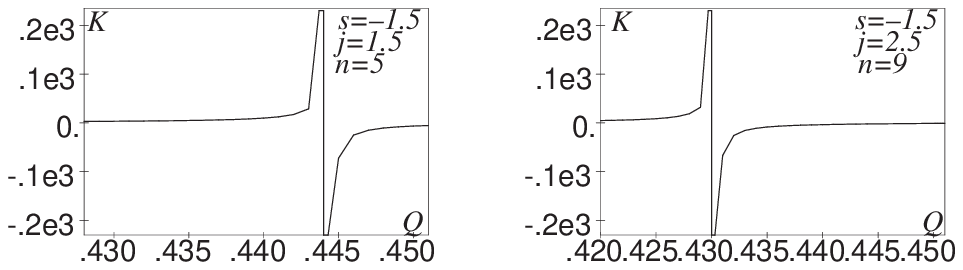}  \\
\caption{\label{fig6} The panels show trajectories of the slope $K=d
\omega_I/d\omega_R$ versus $Q$ near the critical point $Q_{cp}$ for
the first oscillatory QNMs of the scalar, Dirac, and RS fields.}
\end{figure}

To compare them farther more, we take
$K=\frac{d\omega_I}{d\omega_R}$ as the slope of the QNMs. For
$Q<Q_{cp}$, Fig. \ref{fig6} shows that the QNMs have a positive
slope and hence large values of $\omega_R$ correspond to large
values of $\omega_I$. However, for $Q>Q_{cp}$, the slope is negative
and hence large values of $\omega_R$ correspond to small values of
$\omega_I$. At $Q=Q_{cp}$, the slope becomes infinity. The
transition from negative to positive value of $K$  occurs via a
infinite discontinuity, characteristic of a second order phase
transition. The trajectories of the slope of the QNMs take the same
form as that of $-C_{Q}$ versus $Q$.

We all know that there are two characteristic parameters for any
perturbation of a black hole background: the oscillation time scale
$\tau_R=1/\omega_R$ and the damping time scale
$\tau_I=1/|\omega_I|$. Near the critical point, although the
increment of $\tau_I$ increases monotonously as $Q$ increases, the
increment of $\tau_R$ decreases as $Q$ increases for $Q<Q_{cp}$ and
it dose not change at all at $Q=Q_{cp}$, but it increases as $Q$
increases for $Q>Q_{cp}$. It is curious that the change of the
increment of $\tau_R$ presented here is similar to that of the
temperature of the black hole when it takes the same quality of
heat.  If we take a slope of the time scale as
$K_\tau=d\tau_I/d\tau_R$,  the trajectory of the slope takes the
same form as that of $C_{Q}$ versus $Q$ because
$K_\tau=-K(\omega_R/\omega_I)^2$. Above discussions show us that the
critical point of the QNMs may be associated with the SOTPT point.

In summary, we study the relation between the QNMs and the SOTPT for
the RN black hole and find the following results: if the real part
of the quasinormal frequencies arrives at its maximum at the SOTPT
point for given overtone number and angular quantum number, the QNMs
will start to get a spiral-like shape in the complex $\omega$ plane,
and both the real and imaginary parts will become the oscillatory
functions of the charge. The QNMs will (not) take a spiral-like
shape if the angular quantum number or the overtone number larger
(less) than this given value. If a black hole does not possess the
SOTPT point, its QNMs never take a spiral-like  in the complex
$\omega$ plane. Besides the fact that the critical point $Q_{cp}$ of
the QNMs is in good agreement with the SOTPT point $Q_{sp}$, the
transition from negative to positive value of the slope $K$
($K_\tau$) of the QNMs (time scale) occurs via a infinite
discontinuity, and the trajectories of $K$ and $K_\tau$ take the
same form as that of the heat capacity. These facts show that the
critical point of the QNMs may be associated with the SOTPT point
and the quasinormal frequencies carry the thermodynamical
information of the RN black hole.


\hspace*{1.0cm}

[{\bf{Acknowledgments}}] This work was supported by the National
Natural Science Foundation of China under Grant No. 10675045; the
FANEDD under Grant No. 200317; and the Hunan Provincial Natural
Science Foundation of China under grant no. 07JJ3016.


\begin{thebibliography}{99}


\bibitem{Chand75} S. Chandrasekhar and S.  Detweller,  Proc. R.
Soc. Lond. A {\bf  344}, 441  (1975).

\bibitem{Andersson} N. Andersson and H. Onozawa,
Phys. Rev. D  {\bf 54}, 7470 (1996).


\bibitem{Hod} S. Hod,
Phys. Rev. Lett.  {\bf 81}, 4293 (1998).

\bibitem{Dreyer} O. Dreyer,
Phys. Rev. Lett.  {\bf 90}, 081301 (2003).

\bibitem{Maldacena} J. Maldacena, Adv. Theor. Math. Phys.
  {\bf  2},  231 (1998).

\bibitem{Witten} E. Witten,  Adv. Theor. Math. Phys.
  {\bf  2},  253 (1998).

\bibitem{Kalyana} S. Kalyana Rama and B. Sathiapalan, Mod. Phys. Lett.
 A {\bf  14},  2635 (1999).


\bibitem{Davies1}
P. C. W. Davies, Proc. Roy. Soc. Lond. A {\bf 353}, 499 (1977).

\bibitem{Davies2}
P. C. W. Davies,  Rep. Prog. Phys. {\bf 41},  1313 (1978).


\bibitem{Davies3}
P. C. W. Davies,  Class. Quant. Grav. {\bf 6}, 1909 (1989).


\bibitem{Newman} E. Newman and R. Penrose, J. Math. Phys. (N. Y.) {\bf 3}, 566
(1962).



\bibitem{Jing}
Jiliang Jing, Phys. Rev. D {\bf 71}, 124006 (2005).


\bibitem{Khanal} U. Khanal,
Phys. Rev. D  {\bf 28}, 1291 (1983).

\bibitem{Castillo}G. F. Torres del Castillo and G. Silva-Ortigoza,
Phys. Rev. D  {\bf 42}, 4082 (1990).



\bibitem{WS}W. H. Press and S. A. Teukolsky, Astrophys. J. {\bf 185}, 649 (1973).

\bibitem{ET}E. T. Newman and R. Penrose, J. Math. Phys. (N. Y.) {\bf 7}, 863
(1966).

\bibitem{ET-1}J. N. Goldberg, A. J. Macfarlane, E. T. Newman, F. Rohrlich and E.
C. G. Sudarshan, J. Math. Phys. (N. Y.) {\bf 8}, 2155 (1967).


\bibitem{Leaver} E. W. Leaver, Pro. R. Soc. Lond. A {\bf 402},  285
(1985).

\bibitem{Leaver1} E. W. Leaver, Phys. Rev. D {\bf 34}, 384 (1986).




\end{thebibliography}
\end{document}